\documentstyle[epsf,psfig]{mn2e}

\title{First science with SALT: peering at the accreting polar
caps of the eclipsing polar SDSS J015543.40+002807.2 }

\author[D. O'Donoghue et al.]
{D. O'Donoghue$^1$, 
D.A.H. Buckley$^{1,2}$, 
L.A. Balona$^1$, 
D. Bester$^2$, 
L. Botha$^1$, \and
J. Brink$^{1,2}$,
D.B. Carter$^1$,
P.A. Charles$^1$, 
A. Christians$^1$, 
F. Ebrahim$^{1,2}$, \and  
R. Emmerich$^{1,2}$,
W. Esterhuyse$^2$,
G.P. Evans$^1$,
C. Fourie$^1$, 
P. Fourie$^1$, \and  
H. Gajjar$^{1,2}$, 
M. Gordon$^1$, 
C. Gumede$^2$,
M. de Kock$^2$, 
A. Koeslag$^2$, 
W.P. Koorts$^1$, \and  
H. Kriel$^1$, 
F. Marang$^1$,  
J.G. Meiring$^2$,
J.W. Menzies$^1$, 
P. Menzies$^1$, 
D. Metcalfe$^1$, \and  
B. Meyer$^1$, 
L. Nel$^2$, 
J. O'Connor$^1$, 
F. Osman$^1$,
C. du Plessis$^1$, 
H. Rall$^1$, \and  
A. Riddick$^1$, 
E. Romero-Colmenero$^1$, 
S.B. Potter$^{1}$, 
C. Sass$^1$, 
H. Schalekamp$^2$, \and
N. Sessions$^2$,
S. Siyengo$^1$, 
V. Sopela$^1$,
H. Steyn$^1$,  
J. Stoffels$^1$,
J. Stoltz$^1$, \and  
G. Swart$^2$,
A. Swat$^2$, 
J. Swiegers$^2$,
T. Tiheli$^1$,
P. Vaisanen$^1$,
W. Whittaker$^2$, \and  
F. van Wyk$^1$ 
\\
\\
$^{1}$South African Astronomical Observatory, Observatory, 7935, Cape Town, 
South Africa\\ 
$^{2}$Southern African Large Telescope Foundation, Observatory, 7935, 
Cape Town, South Africa\\
}

\date{}

\date{Accepted \ \ \ \ \ \ \ \ \ \ \ \ \ \ \ \ \ \ \ \ \ \ \ \ \ \ \ \ \ \ \ Received}

\begin{document}

\maketitle

\begin{abstract}
We describe briefly the properties of the recently completed Southern
African Large Telescope (SALT), along with its first light imager
SALTICAM. Using this instrument, we present 4.3 hr of high speed unfiltered
photometric observations of the eclipsing polar SDSSJ015543.40+002807.2 with 
time resolution as short as 112 ms, the highest quality observations of this 
kind of any polar to date. The system was observed during its high
luminosity state.  Two accreting poles are clearly seen in the 
eclipse light curve. The binary system parameters have been constrained: 
the white dwarf mass is at the low end of the range expected for cataclysmic
variables. Correlations between the positions of the accretion regions
on or near the surface of the white dwarf and the binary system parameters 
were established. The sizes of the accretion regions and their relative 
movement from eclipse to eclipse were estimated: they are typically $4^{\rm o}$
-7$^{\rm o}$ depending on the mass of the white dwarf. The potential of these
observations will only fully be realised when low state data of the same
kind are obtained and the contact phases of the eclipse of the white
dwarf are measured.
\end{abstract}


\section{Introduction To SALT}
\label{salt}

The Southern African Large Telescope, colloquially known by its acronym SALT, 
has recently been completed. It is now in the final stages of commissioning 
and first stages of science operations. This paper describes the properties 
of the telescope and its imaging camera SALTICAM. It then presents the first 
science observations of an eclipsing magnetic cataclysmic variable, SDSS 
J015543.40+002807.2, analyses the data and shows how our knowledge of this
star has been advanced.

Although SALT has been described in the proceedings of SPIE conferences 
(e.g. Stobie et al. 2000; Meiring et al. 2003; Meiring \& Buckley 2004), 
these publications are not available in all astronomical libraries and not 
accessible on the Internet. Thus, a brief description of SALT 
is appropriate. Its basic principle of operation is similar to that of the 
Arecibo radio telescope, employing a spherical primary mirror and a payload 
carrying the instrumentation which tracks the spherically-shaped focal surface. 
This surface is concentric with the primary mirror and located at a distance 
of 13.08 m (half the radius of curvature) from the primary mirror. The overall
optical design of SALT is described in Swat et al. (2003).

The primary mirror is tilted so that its axis of symmetry is fixed at an
inclination of 37$^{\rm o}$ to the zenith. The telescope can move in azimuth, 
but not in elevation so it can observe any celestial object which reaches an 
altitude of 53$^{\rm o}$. The tracking capability described below extends this 
range of altitudes from 47$^{\rm o}$ to 59$^{\rm o}$.

The basic concept for this telescope was proposed by Ramsey \& Weedman (1984) 
and its first implementation was the construction of the Hobby-Eberly Telescope
at McDonald Observatory (HET). SALT is the second such telescope and started 
off its development process using the design of the HET. However, many aspects 
of this design were improved upon. 

\subsection{Primary mirror}

The primary mirror is comprised of 91 hexagonal spherical segments 1 m across, 
50 mm thick, and made of Astro-Sitall (Swiegers \& Gajjar 2004). The arrangement
of the primary segments is also hexagonal such that a circle of diameter 11 m 
would pass through the vertices of the hexagon.  The individual segments are 
shaped to a global radius of curvature of the primary of $26.165\pm0.0005$ m, 
and figured to better than 1/15th wave rms. Each segment is supported on a 
mirror mount with nine points of support; three actuators move the segment in 
tip/tilt and piston. The mirror mounts are ``plugged" in to a steel space-frame 
truss which itself is kinematically supported on the telescope structure. 

Alignment of the mirror segments with respect to each other to an accuracy of
0.06" rms takes place in the evening twilight and is achieved using a 
Shack-Hartmann wavefront camera, located at the centre of curvature of the 
primary mirror in a tower alongside the telescope building. The alignment is 
maintained using a system of capacitive edge sensors which are bonded to the 
six sides of each mirror segment, generating tiny but detectable changes in 
capacitance as the segments move with respect to each other during the night. 
These signals are fed to a closed-loop control system which adjusts the segment 
positioning every 20 s to maintain the original alignment of the mirror.

\subsection{Tracker and Payload}

The payload at the prime focus is positioned using a ``Tracker" which is 
capable of movement in x, y, z, tip, tilt and rotation with accuracies of 
$\sim5$ microns and 1 arcsec (despite carrying a weight of more than a metric 
ton!). Tracking celestial objects requires adjustment of x and y, along with 
concomitant changes in tip and tilt. The focus of the telescope is controlled 
by movement in z. The payload is also rotated about the telescope optical
axis during observation to remove field rotation at the focal plane. 

The payload includes a 4-mirror spherical aberration corrector (SAC) 
(O'Donoghue 2000) to correct the huge spherical aberration of the primary 
(the circle of least confusion of the uncorrected prime focus is about
20 arcmin in diameter) and feed the corrected prime focus with an f/4.2 beam. 
The SAC yields a circular, flat, science field of view of 8 arcmin in diameter, 
with a 1 arcmin annulus around it for guide stars used in closed loop guiding.

The entrance pupil of the SAC is 11 m in diameter and can thus accept light 
from celestial targets from the entire primary mirror. However, as the Tracker 
tracks celestial objects, the alignment of the SAC entrance pupil must 
necessarily be offset with respect to the primary mirror, so that parts of the 
primary mirror outside the entrance pupil do not contribute to light reaching 
the focal plane. Fortunately, the exit pupil of the optical system is 
readily accessible and a moving hexagonal baffle at this position is kept
aligned with the primary mirror, thus preventing stray light from the 
periphery of the primary mirror reaching the focal plane. The telescope is 
equipped with an atmospheric dispersion compensator (O'Donoghue 2002) to 
enable access to wavelengths as short as 320 nm without image quality 
degradation arising from atmospheric dispersion. 

As mentioned, the axis of symmetry of the primary mirror is tilted with respect 
to the zenith by 37$^{\rm o}$, so that the accessible declination range is 
+10.5$^{\rm o}$ to -75.3$^{\rm o}$. Celestial targets are accessible to the 
telescope when they enter an annular region which is centred on the zenith and 
has an angular distance (with respect to the zenith) of 37$^{\rm o}$. The range 
of the Tracker is $\pm6^{\rm o}$ so that a celestial target is accessible when 
its zenith distance lies in the range 31$^{\rm o}$ to 43$^{\rm o}$. These 
constraints permit objects to be tracked for a duration of 1 hr at the northern 
end of the accessible declination range, and more than 3 hr at the southern 
end.

The mode of operation of the telescope is straightforward: the telescope 
is slewed to the appropriate azimuth for a given celestial target. The primary 
mirror is stationary at this azimuth during the observation. When the desired 
celestial target enters the annular region of accessibility, it is tracked 
using the Tracker. Most celestial targets are accessible twice in a given 
night: once in the eastern sky and once in the western sky. At the extreme 
north and south ends of the declination range, targets can be re-acquired 
by re-positioning of the telescope in azimuth, thus permitting more than 
two pointings.

\section{SALTICAM: The SALT Imaging Camera}
\label{salticam}  

Among the First Light instrumentation for SALT (Buckley et al. 2003, 2004)
is a UV-visible simple imager, SALTICAM.

Due to the interests of many of the scientists in the consortium partners, 
capability at the shortest wavelengths accessible from the ground ($\sim320$ nm)
and high time resolution in the detectors are features of the first generation 
of instrumentation on the telescope. The observations presented in this paper 
were made with SALTICAM. This instrument has already been described elsewhere 
(O'Donoghue et al.  2003) but a brief summary is given here to provide context 
for the data to be discussed below.

SALTICAM was designed to fulfill a number of functions: a camera to provide 
an instrument for the telescope as early during its commissioning and testing
phase as possible; the acquisition camera for the telescope; and a simple 
imaging instrument with a high time resolution capability. Its detector 
comprises a mosaiced pair of E2V Technologies 44-82 CCDs: 2048 x 4102 x 15 
micron pixels (60 x 30 mm$^2$ of imaging area). Each CCD is coated with the E2V 
Astro Broad Band coating providing excellent UV sensitivity, is made from deep 
depletion silicon to provide additional sensitivity and less fringing in the 
I band, and has two low-noise ($\sim3$ electron/pix) readout amplifiers. 
Each CCD has independent vertical clock lines for each half of the chip and can 
thus be operated in Frame Transfer mode.

The f/4.2 focal ratio of SALT yields a plate scale of 224 microns/arcsec so 
that an 8 arcmin field spans 107.5 mm with considerable oversampling of 15 
micron pixels. SALTICAM was thus equipped with a 7-element lens system to 
re-image the 8 arcmin field on the pair of CCDs described above at a scale of 
107 microns/arcsec (and a focal ratio of f/2.0). The lenses are made of UV 
transmitting crystals: fused silica, BaF$_2$ and CaF$_2$ thereby affording 
excellent transmission for wavelengths as short as 320 nm.

As mentioned, high time resolution capability was one of the
scientific drivers for the instrument, with 0.1 second sampling for
point sources identified as a requirement. The large CCD 44-82s have
considerable on-chip capacitance, necessitating a rather slow vertical
clocking rate of 100 microseconds/row.  Vertical clocking of half of
each chip in standard frame transfer operation would thus add
$\sim0.2$ seconds, thereby ruling out achievement of the requirement.
To circumvent this difficulty, ``slot mode" operation was devised. In
this mode, a mask with a slot of width projecting to 144 (unbinned)
rows (20 arcsec) on the CCDs is positioned just above the frame
transfer boundary of the chips (which are aligned co-linearly with
respect to each other). The slot does not lie on the CCDs, of course,
but $\sim2$ mm above, so there is a region of partial vignetting. Only
the central 72 rows receive all the light. The slot extends the full 8
arcmin in the horizontal direction. After each exposure, only 144 rows
are transferred across the frame transfer boundary, thereby shortening
the vertical clocking overhead to 14 millisec. Each image takes a few
exposures to ``migrate" to the readout register but this presents no
particular problem and allowance is made for this in the time-stamping
of the data which is done by hardware interrupts from the SAAO time
service.  This time service is accurate to better than 1ms.

The detectors are housed in an evacuated cryostat, cooled to 160 K by a 
CryoTiger closed-cycle cooler, and controlled by an SDSU II CCD controller with 
custom developed software. A standard 100 mm Prontor shutter was modified to 
use 2 solenoids for ``flip-flop" opening and closing (to avoid needing a 
solenoid to be powered up and therefore a potential source of heat for the 
duration of each exposure). The shutter is, of course, held open 
during rapid sampling as described in the previous paragraph. Finally, a 
filter unit allows one of 8 exchangeable, broad band filters to be inserted 
into the beam under software control for any given exposure.

Further description of the instrument, including quantitative details and
more information on the high time resolution capability, is given in 
O'Donoghue et al. (2003, 2004).

\begin{table*}
\caption{High speed clear filter photometry and eclipse timings of SDSS 
J015543+002807. Contact times (all in days) listed in the table refer to the 
ingress/egress of Spot 1 (first line) and Spot 2 (second line) of the 
corresponding run. Missing entries arise because the timing was not available, 
due to gaps in the data. The O-C values (in parentheses) for mid-eclipse 
timings are in units of 0.000001d. See text for more details.  
{\label{tab:observations}}}
\begin{tabular}{@{}cccccccccccc}
Date &  Exp. &    Run   &Eclipse&          &  First    &   Mid       & Second      & Mid         &  Third  &  Mid   & Fourth  \\
2005 &  time &  length  & cycle &          &contact    & ingress     & contact     & eclipse     & contact & egress & contact \\
     & (ms) & (min)   &       & HJD      &           &             &             & (and O-C)       &         &        &         \\
Aug\\    
    05/06 & 285 & 10.1 & 10234 & 2453588+ &            &             &             &             & 0.648009    & 0.648021    & 0.648036\\
          &     &      &       & 2453588+ & 0.644692   & 0.644705    & 0.644715    &             &             &             &         \\
    06/07 & 230 & 12.3 & 10250 & 2453589+ & 0.612659   & 0.612670    &             & 0.614473(-4)& 0.616269    & 0.616277    & 0.616288\\
          &     &      &       & 2453589+ & 0.612930   & 0.612943    & 0.612954    & 0.614474(-4)& 0.615998    & 0.616005    & 0.616025\\
    06/07 & 230 & 33.4 & 10251 & 2453589+ & 0.673189   & 0.673200    & 0.673211    & 0.674992(-2)& 0.676780    & 0.676785    & 0.676795\\
          &     &      &       & 2453589+ &            & 0.673468    & 0.673476    & 0.674992(-2)& 0.676504    & 0.676517    & 0.676533\\
    10/11 & 230 & 26.3 & 10317 & 2453593+ & 0.667264   & 0.667277    & 0.667293    & 0.669069(-3)& 0.670848    & 0.670862    & 0.670872\\
          &     &      &       & 2453593+ & 0.667541   & 0.667552    & 0.667562    &             &             &             &         \\
    11/12 & 230 & 26.5 & 10332 & 2453594+ & 0.575008   & 0.575019    & 0.575029    &             &             &             &         \\
          &     &      &       & 2453594+ & 0.575270   & 0.575283    & 0.575296    & 0.576817(0) & 0.578341    & 0.578351    & 0.578365\\
Sep\\
    06/07 & 112 & 45.4 & 10761 & 2453620+ & 0.536510   & 0.536518    & 0.536529    & 0.538317(-6)& 0.540104    & 0.540115    & 0.540126\\
          &     &      &       & 2453620+ & 0.536782   & 0.536794    & 0.536800    & 0.538318(-5)& 0.539837    & 0.539843    & 0.539863\\
    06/07 & 112 & 15.5 \\
    06/07 & 112 & 45.9 & 10762 & 2453620+ & 0.597027   & 0.597035    & 0.597053    & 0.598832(-7)& 0.600602    & 0.600630    & 0.600638\\
          &     &      &       & 2453620+ & 0.597300   & 0.597309    & 0.597316    & 0.598834(-5)& 0.600353    & 0.600360    & 0.600376\\
    07/08 & 112 & 41.5 & 10777 & 2453621+ & 0.504795   & 0.504804    & 0.504819    & 0.506600(+15)&0.508376    & 0.508396    & 0.508419\\
          &     &      &       & 2453621+ & 0.505068   & 0.505079    & 0.505095    & 0.506604(+19)&0.508116    & 0.508128    & 0.508144\\
\end{tabular}
\end{table*}

\section{Eclipsing AM Her Stars (or Polars)}
\label{polars}

Amongst the shortest period binaries known are AM Her stars, also called 
Polars. A subclass of cataclysmic variables, they are semi-detached binaries 
in which a Roche lobe-filling lower main sequence star (usually an M star) 
transfers mass to a highly magnetized ($\sim10^7$ to 10$^{8.5}$ Gauss) white 
dwarf. The magnetic field of the white dwarf (the primary) inhibits the 
formation of the standard thin accretion disc found in most cataclysmic 
variables and instead channels the gas flowing from the lower main sequence 
star (the secondary) to accretion regions at or near the magnetic poles of 
the white dwarf. About 85 such systems are known with orbital periods 
ranging from 87 min to $\sim$8 hr (see Schwope et al. 2004 for a review
of eclipsing polars, and Wickramasinghe \& Ferrario 2000, Warner 1995
and Cropper 1990 for more general reviews of AM Her stars).

In the broader context, these systems are valuable laboratories for
studying magnetic accretion on to compact objects, found in many other 
astrophysical scenarios from pulsars to active galactic nuclei.

Usually, the gas falling on to the white dwarf is thermalized in a strong 
shock at or above the white dwarf surface and may outshine most other sources 
of radiation in the system, emitting radiation from X-ray to infrared 
wavelengths. The footprint on the white dwarf where the accretion occurs 
is also very small: typically 10$^{-4}$ of the surface area of the star.

Eclipsing polars are very valuable examples of these systems as the eclipse 
of the white dwarf by the secondary allows indirect mapping of the gas flow 
and accretion on to the white dwarf. High time resolution is, however, 
critical as the accretion regions are eclipsed by the secondary in only 1-6 s. 
SALTICAM's efficiency at all optical wavelengths and high time resolution 
makes it highly suitable for studies of eclipsing polars such as SDSS 
J015543+002807, the topic of this paper. The primary aim is to resolve the 
eclipse ingress/egresses of the accretion region(s). Similar work on other
eclipsing polars using a Superconducting Tunnel Junction detector has been 
described by Perryman et al. (2001), Bridge et al. (2002) and Reynolds et al. 
(2005). There is also discussion of unpublished results obtained by ULTRACAM 
and OPTIMA in Schwope et al. (2004).

\section{SDSS J015543+002807}

This object was identified in the Sloan Digital Sky Survey as an
eclipsing cataclysmic variable (Szkody et al. 2002). Follow-up optical
photometry, showing very deep eclipses, has been published by Dubkova et 
al. (2003) and Woudt et al. (2004). 

Wiehahn et al. (2004) (hereafter W2004) confirmed that the star was indeed 
a polar on the basis of polarimetry which showed variations in circular 
polarization between -10 and +10 per cent, along with strong linear 
polarization.  The photometry obtained as part of the polarimetric data 
showed that the system brightness declined very rapidly (in a matter of a 
few seconds or even less).

During the W2004 observations the system was in a low luminosity 
state. Schmidt et al. (2005) (hereafter S2005) confirmed the polarization, 
extending the range of measurements to the ``high" brightness state of the 
system. They also discussed X-ray observations and found that the optical
emission lines changed to absorption lines during certain orbital phases. 
Neither W2004 nor S2005 could decide whether or not there were one or two 
accreting poles in the system.

Because of the rapid, deep eclipses, implying that the optical emission 
originates from a compact region on the white dwarf, SDSS J015543+002807
is a highly suitable target for observations with SALTICAM as described
below.

\begin{figure*}
\epsfxsize=15.6cm
\epsffile{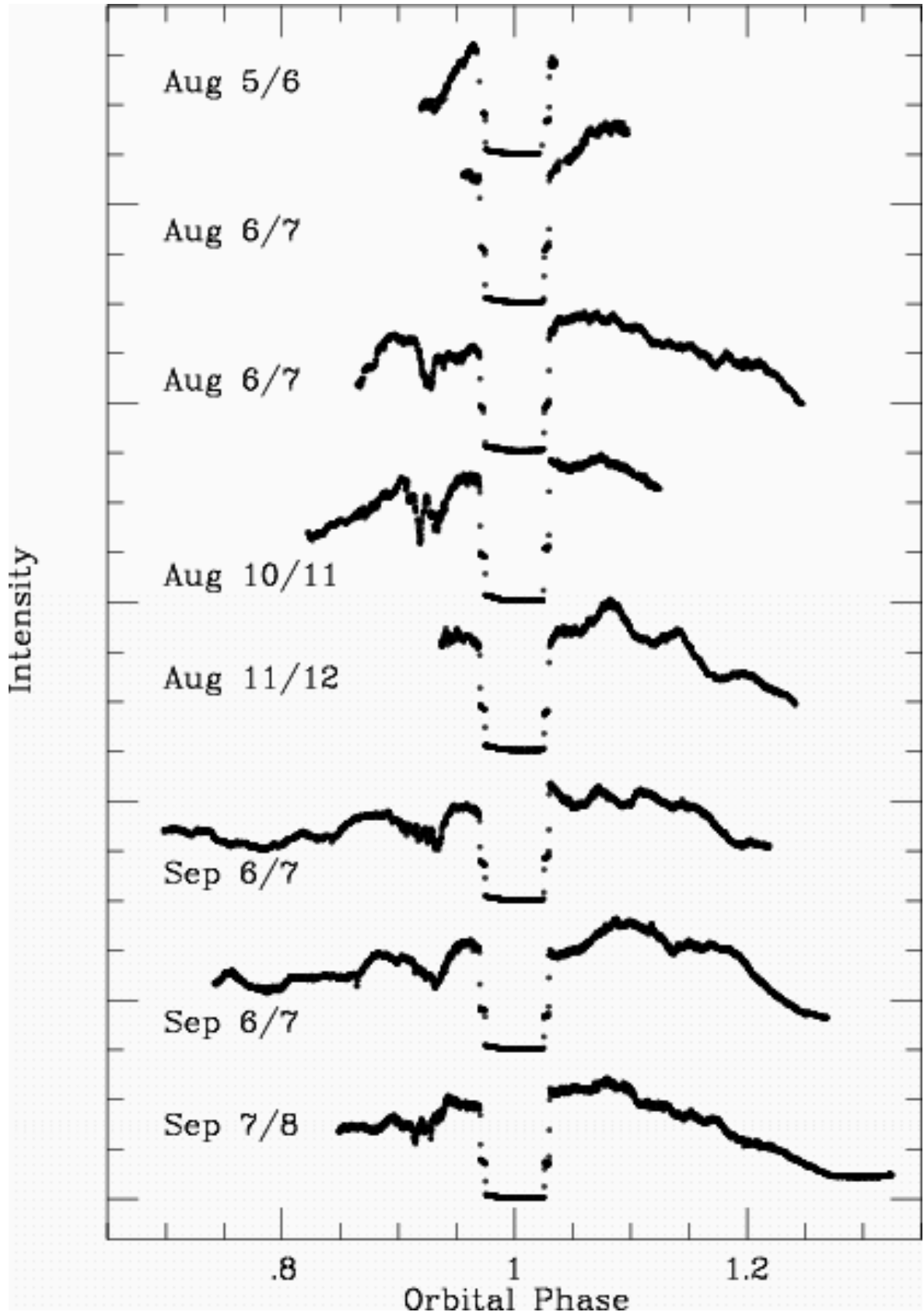}     
\caption{High speed photometry listed in Table 1 (except for the second
section on 2005 Sep 06/07). The horizontal axis is orbital phase using
the ephemeris of Equation~1. Its carets (0.05 phase) correspond to 
$\sim260$ sec. The vertical axis is relative intensity in arbitrary
units (i.e. linearly proportional to the brightness of the star). Its 
zero point is the lowest caret on the vertical axis. Individual runs have 
been offset from adjacent runs by 3 carets on the vertical axis. The points 
are 1-sec averages of the original data.}
\label{fig1}       
\end{figure*}

\section{Observations}

During the SALTICAM commissioning observing run, time series imaging   
of SDSS J015543+002807 was obtained by SALT+SALTICAM on the nights of 
2005 August 5/6, 6/7, 10/11, 11/12 and 2005 September 6/7 and 7/8. 
Exposure times ranged between 112 and 285 millisec (the time to move
each image behind the mask, during which image smearing takes place,
is 14 $\mu$s). A summary log of 
the 4.3 hr of observations appears in Table~\ref{tab:observations}.
Altogether 9 time series were obtained, 8 of which included eclipses 
of the target star: the second sequence on 2005 September 06/07 did 
not include an eclipse. 

The observations were made with the instrument in ``slot mode" as
described in Section \ref{salticam}. There were problems in managing the 
data flow between the SDSU II CCD controller and PC during the August 
observations, so gaps of about 6 sec, while the data were being displayed 
and written to disk, appear in these data at intervals of typically 
18-20 sec.  These problems had been almost entirely solved by the time 
of the September data. The instrument maintained strict time using a 
stable 1 kHz signal originating from a GPS receiver which also provided
UTC. The instrument computer stamped the header of each frame with the 
time of the exposure and this was subsequently used to calculate the
Heliocentric Julian Date (HJD) for all frames. We have not used the
more uniform Barycentric Julian Dynamical Time as previous timings have
not used this system. To convert all timings in this paper to BJDD
(JD(TDB)), add 67.12 s to the August data and 67.14 s to the September 
data.

Because of the shortness of the exposure times, and the decline of
the star to fainter than 20th magnitude at mid-eclipse,
the observations were made through a clear filter comprising 5 mm of
WG295: the filters in SALTICAM are located in the strongly converging f/2
beam so having no filter would significantly degrade image quality.
BVR filtered observations were tried but the loss of light, combined with 
poor image quality at the time these data were obtained, has rendered
them of limited use.

\begin{figure*}
\epsfxsize=16.4cm
\epsffile{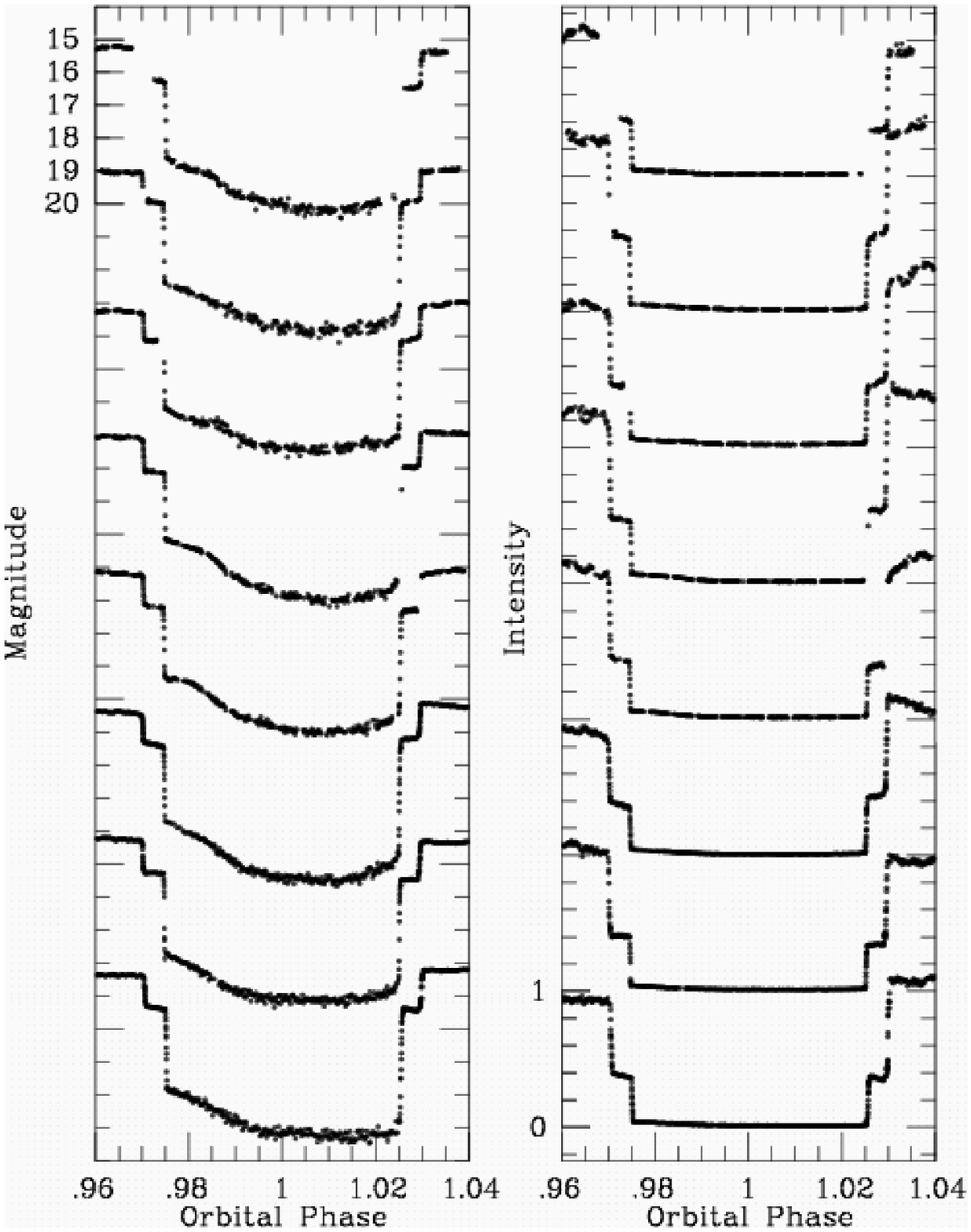}     
\caption{High speed photometry shown in Fig. 1 around the time of eclipse
with the vertical axis in magnitudes (left panel) and intensity (right panel). 
The ordering of the individual runs is as in Fig. 1. The horizontal axis is 
orbital phase using the ephemeris of Equation~1. Individual carets (0.005 
phase) correspond to $\sim26$ sec. The vertical axis of the left panel is 
in magnitudes and is intended to show detail at the bottom of eclipse. Its 
overall zero point is accurate to $\sim0.3$ mag (the relative zero point from
run to run is known much more accurately). Individual runs in this panel have 
been offset from adjacent runs by 4 carets on the vertical axis. The vertical 
axis of the right hand panel is intensity as in Fig. 1. The zero level of
each run occurs at the large caret marks (i.e. individual runs have been 
offset from adjacent runs by 5 carets on the vertical axis). With the exception 
of the points occurring in the precipitous brightness changes, where the 
original time resolution has been retained, the points in both panels are 
1-sec averages of the original data.}
\label{fig2}       
\end{figure*}

The instrument thus provided a sequence of images 8 arcmin long, but
only 20 arcsec wide. These images were split into 4 distinct sections, 
corresponding to the 4 readout amplifiers of the two CCDs. In order
to facilitate the shortest possible exposures, $4\times4$ (August) and
$6\times6$ (September) prebinning was used. This did not lead to undersampling 
of the images: each prebinned pixel corresponded to 0.56 (August) and 0.84 
(September) arcsec and the combination of seeing and the image quality of 
the telescope at the time was modest, between 1.5 and 2.0 arcsec (FWHM).

High accuracy {\em absolute} photometry with SALT is very difficult because
of the moving entrance pupil and resulting variable aperture as described 
in Section \ref{salt}. Fortunately, all stars within the field of view
of the telescope lose the same fraction of light due to mismatch of the
pupil and the primary mirror (as well as the small gaps between the primary
mirror segments). Provided the brightness of the comparison star(s) is known, 
the absolute brightness of other stars in the field can be recovered. The
data presented here are all referenced to the nearby star, 27 arcsec to the 
west and 5 arcsec north of the target star; its magnitudes and colours in the 
Sloan Digital Sky Survey filter set appear in Dubkova et al. (2003). The 
comparison star and the target star were imaged within the same amplifier.
 
No flat field calibration frames were obtained (the calibration system
was not available at the time of the observations) and the frames were simply
bias-subtracted using overscanned pixels from each row. Despite this, the
errors in the photometry are very small ($\sim 0.02$ mag outside of eclipse). 
Stellar magnitudes were extracted using standard CCD photometry techniques 
and the brightness of the target star was referenced to that of the comparison 
star mentioned above.

\section{Light Curves}

Although the August data quality is somewhat affected by the gaps, 
these and (especially) the September data constitute the highest time 
resolution/highest signal-to-noise light curves of eclipsing polars of 
which we are aware. It is thus worth describing them in some detail.

Fig.~\ref{fig1} shows the full data set on an intensity scale (i.e. the
intensity is proportional to the brightness of the target), and phased
using the orbital ephemeris derived in the next section (see 
Equation~1 below).  The light curves show the features
already reported by other authors: an asymmetric orbital modulation 
with a gradual rise to a maximum at orbital phase $\sim0.1$ followed
by a more rapid decline; the eclipse interrupts the light curve just
before maximum. During the orbital phase interval 0.90--0.95 there is
a pre-eclipse dip of variable shape and depth, which has been attributed
to obscuration by the mass transfer stream of the strong emission 
originating at the accretion columns/spots on or close to the white 
dwarf primary. Note that the intensity of each run is on the {\em same}
scale, as the data were referenced to the comparison star mentioned above.

Superimposed on the orbital modulation are stochastic brightness variations
typical of cataclysmic variables. In contrast to the high frequency
flickering in polars such as VV Pup, for example, the variations in this
star occur on a time scale of minutes rather than seconds. On short time
scales, the light curve is quite smooth.

Fig.~\ref{fig2} shows, in the left and right panels, more detail around 
the eclipse in both magnitudes and intensity, respectively, as different 
information is conveyed in each panel.  In order 
to improve the signal-to-noise, especially at the bottom of the eclipse, 
most of the data in Fig.~\ref{fig2} have been averaged: with the exception 
of the portions during the precipitous brightness changes, where the original 
time resolution has been retained (Table~\ref{tab:observations}), the points 
in the plot are 1-sec averages of the original data. 

The most notable feature of Fig.~\ref{fig2} is the stepped shape of the
eclipses, visible in every eclipse (taking into account the missing sections 
in the August data). At phase $\sim0.97$ there is a very rapid decline of 
$\sim1.0$ mag, followed at phase $\sim0.975$ by another very rapid decline 
of $\sim2.5$ mag. There is then a much more gradual further decline of 
$\sim1.5$ mag over a time interval of $\sim105$ sec up until phase 0.0. 
In some light curves, the bottom of eclipse is flat; in others there is
a slight rise ($\sim0.5$ mag) just before the very rapid increase in 
brightness at $\sim1.026$. There is then a plateau similar to that seen
on ingress and lasting $\sim0.005$, followed by the last very rapid 
emergence occurring at $\sim1.030$. 

The right panels show the same data on an intensity scale which 
emphasizes different features to the left panel: the first drop/last
rise is brighter than the second (contrary to the impression given
in the left panel). Also, the fading between phase 0.975 and 1.000 is
much less obvious and the light curve from phase 1.000 to 1.030 looks
essentially flat. {\em The two precipitous brightness changes thus 
account for almost all the flux in these light curves}.

As mentioned already, SDSS J015543+002807 has been established to be
a polar (W2004, S2005). Within this context, Figs.~\ref{fig1}$-$\ref{fig2} 
allow some immediate conclusions to be drawn. These are most easily
appreciated with the aid of Fig.~\ref{fig7}(left). The Roche lobe-filling
M dwarf is shown in black to indicate its minimal contribution to the
light curve. Its limb has just begun to eclipse the photosphere of the
white dwarf. On (or near) the surface of the white dwarf are 2 bright spots
(shown as white in the inset) which give rise to almost all the light 
from the system and which cause the stepped structure of the eclipse
comprising two precipitous brightness changes in a very short time,
with a standstill in between. We will refer to these spots as ``Spot 1"
and ``Spot 2" as indicated in Fig.~\ref{fig7}. We are aware that 
although we have described the features as located on the surface of
the white dwarf, they may, perhaps even more likely, be accretion columns 
{\em above} the surface of the white dwarf. In order to avoid clumsy wording, 
we shall simply refer to them as spots, but all the time bearing in mind 
that they may not be located on the surface of the white dwarf.

The rest of the white dwarf is shown in dark grey: the standstills at
orbital phases 0.9725 and 1.0275 are not entirely flat and usually
show a small decline/rise on ingress/egress indicating that the white
dwarf photosphere, which is being covered/uncovered at these phases
(see Fig.\ref{fig7}), does contribute a small amount of light.

While the shading chosen for the components in Fig.~\ref{fig7} is
intended to illustrate their relative light contributions, and is
somewhat arbitrary, the dimensions are accurately computed from Roche
lobe geometry for a specific mass ratio, inclination and white dwarf 
radius. The location and extent of the spots shown is consistent with 
producing the observed light curves. The relevant parameters are not 
unique, of course, and further analysis will be given in Sections 8 \& 9. 
A similar diagram is shown as fig. 4 in Perryman et al. (2001) in their 
study of the eclipsing polar UZ For which also has a similarly shaped 
eclipse indicative of two bright regions on the surface of the white 
dwarf. Other sources of light in that system make a significant
contribution, however, so the eclipse is not nearly as deep as in
SDSS J015543+002807.

\begin{figure}
\epsfxsize=8.4cm
\epsffile{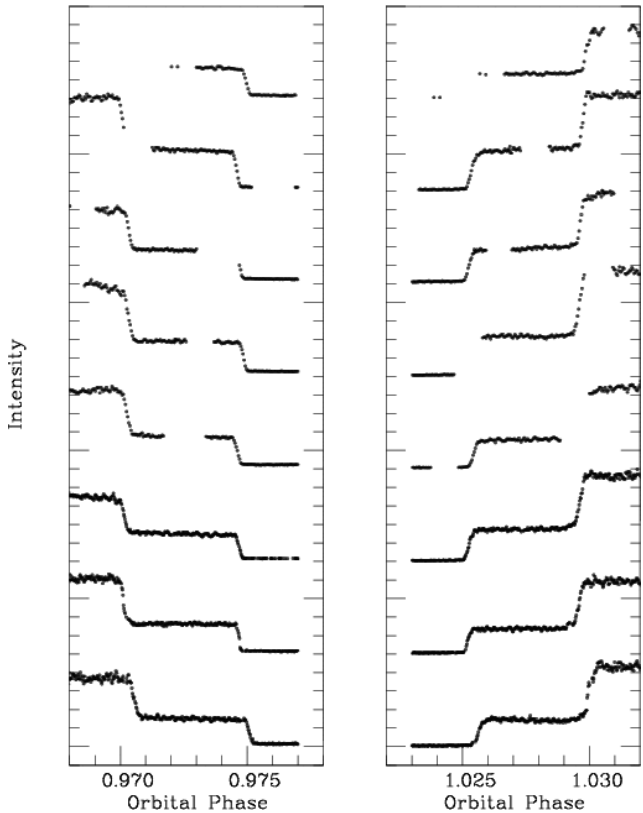}     
\caption{Ingress and egress detail. The ordering of the individual 
runs is as in Fig. 1 and 2. The horizontal axis is orbital phase using 
the ephemeris of Equation~1. Individual carets (0.001 phase) correspond 
to $\sim5.2$ sec. The vertical axis is relative intensity as for 
Fig.~\ref{fig1}. Its zero point is indicated by the lowest 
caret on the vertical axis. Individual runs have been offset from their 
neighbours by 4 carets on the vertical axis.}
\label{fig3}       
\end{figure}

Other aspects of the eclipse light curves shown in Fig.~\ref{fig2} are 
also interesting: the magnitude scale of the left panel was calculated 
by assuming that the comparison star has a ``clear filter" magnitude 
(i.e. weighted over the 600 nm range of sensitivity of the CCDs) of 16.7. 
We expect that this rough zero point will be accurate to $\sim$0.3 mag. 
The resulting brightness of the target is slightly fainter than 15th 
mag just before eclipse. It is thus clear that the target was in its bright 
luminosity state for the duration of the observations presented here (see 
figs. 2-4 of S2005). With respect to its brightness just before or just 
after eclipse, the star is $\sim5$ mag fainter at the faintest part of the 
eclipse.

In other eclipsing polars (e.g. UZ For: Perryman et al. 2001; HU Aqr:
Bridge et al. 2002), there is also a gradual decline in brightness similar 
to that seen in Fig.~\ref{fig2} between orbital phase 0.975 and 0.0.
This is explained as eclipse of emission from the accretion stream as
it flows towards the accretion region(s) on or near the white dwarf.
Clearly a similar explanation is warranted here. Presumably re-appearance
of parts of the stream accounts for the rounded shape to the bottom of
the eclipse as opposed to, say, light contributed by the secondary star. 
We defer detailed study of the accretion stream to a later paper.

\section{Eclipse Ephemeris}

Fig.~\ref{fig3} shows the eclipses of Spots 1 and 2 on an expanded
horizontal scale and at the time resolution of the original data. 
The ingress/egress of the spots takes place in $\sim1-2$ sec. 

\begin{figure}
\epsfxsize=7cm
\epsffile{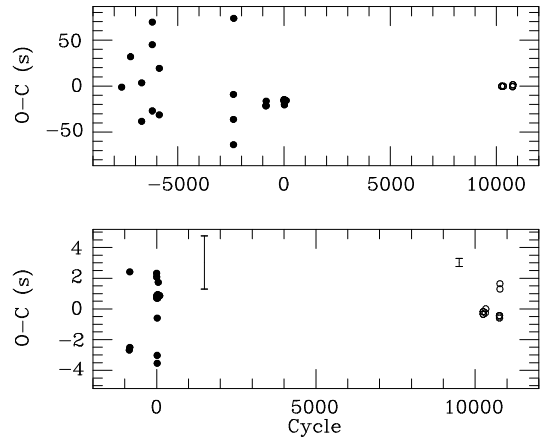}     
\caption{O-C diagram for all available eclipse timings with respect
to the ephemeris of Equation~1. The top panel shows all 
available timings including those listed in S2005 excepting for the two 
with the largest error bars. The ephemeris was derived from the timings 
in the lower panel which are the best quality subset from the upper panel. 
The error bar positioned at cycle 9500, O-C 3.0 s in the lower panel 
shows the 2-$\sigma$ uncertainties for the SALTICAM timings. The error bar 
positioned at cycle 1500, O-C 3.0 s shows the 2-$\sigma$ uncertainties 
for the W2004 timings (see their table 2). }
\label{fig4}       
\end{figure}

Estimates of the time of first to fourth contact as well as mid
ingress and mid egress (defined as the time when half the light in
the spot is eclipsed) of both spots were made by inspecting large scale 
plots and referring to the times of the points on the plot. These times
are listed in Table~\ref{tab:observations}, with the values for Spot 1
on the first line and those for Spot 2 on the second line of the entry
for that run. We estimate that the 1-$\sigma$ random error of this method 
is 1 sampling time (second column of Table~\ref{tab:observations}). It
turns out that the time of mid-eclipse (defined as the mean of the times
of mid-ingress and mid-egress) of Spot 1 is almost identical to that of
Spot 2. Accordingly, a separate timing of mid-eclipse for each spot was 
made and the results are also listed in Table~\ref{tab:observations} where the cycle
numbers use the zero point of W2004. These timings were used along with 
all those listed in S2005 and W2004 to derive a linear ephemeris for mid 
eclipse; the residuals with respect to this ephemeris are shown in the top 
panel in Fig.~\ref{fig4}. The large scatter at negative cycle numbers is due
to these timings being made from conventional CCD observations on modest
sized telescopes with exposure times of minutes. The fast photometric
and polarimetric data of Woudt et al. (2004), W2004 and the present data
(cycles -1000 to 11000) have scatter lower by at least a factor of 10
so these were selected and used to derive the ephemeris for mid-eclipse
of both spots shown in Equation~1. 
\vspace{3mm}
\begin{tabular}{cll}
{\rm HJD} & = & $2452969.322083 + 0^{d}.0605163312$ {\rm E}\ \ \ \,(1)\\
{\rm  Mid Ecl}   &   & \,\ \ \ \ \ \ \ \ \ \ \ \ \ \ \ \ $\pm$5  \ \ \ \ \ \ \ \ \ \ \ \ \ \ \ \ \ \,$\pm7$ 
\end{tabular}   
\vspace{2.5mm}

\noindent
The residuals with respect to this ephemeris are plotted in the lower 
panel of Fig.~\ref{fig4} and listed, for the SALTICAM data, in parentheses 
in the column in Table~\ref{tab:observations} labelled ``Mid eclipse". The
ephemeris is consistent with previous ephemerides (e.g. W2004, S2005), 
extends over a much longer baseline and is much higher quality. It was used 
for the conversion from time to orbital phase throughout this paper. Note
that the epoch in the ephemeris is {\em not} necessarily that of
conjunction of the two stars; the white dwarf was not detected in our
observations.

With the exception of the last SALTICAM run on 2005 Sep 7/8, the scatter 
of the SALTICAM residuals is less than 0.5 s (peak to peak) and consistent 
with measurement error (see the caption to Fig. 5 for an explanation of
the measurement error). The timings associated with the last run differ 
from the ephemeris by more than measurement error: this is obvious from
Fig.~\ref{fig3} in which the lowest eclipse shown occurs distinctly 
later than in the other runs. Post facto checks of the instrument timings 
give no reason to doubt the accuracy of the timings from the last run. 
We believe this is a real difference.  

The residuals for the times of mid eclipse of Spot 1 and 2, which are 
entirely independent measurements, are very similar and also much smaller 
than the scatter of all the O-C values in Table~\ref{tab:observations}. This 
not only shows the simultaneity of mid eclipse for Spots 1 and 2 as 
previously mentioned, but also shows that there is intrinsic scatter in 
the times which is larger than measurement error.  The former result also 
confirms the identification of the ingress/egress of Spot 1 as the first/last 
precipitous brightness change and that of Spot 2 as the second/third 
precipitous brightness change; further evidence is the similar size of the 
corresponding brightness changes. The simultaneity of mid eclipse times of 
the two spots is remarkable: the largest difference in 
Table~\ref{tab:observations} is 0.3 sec.

\section{Binary Parameters}

The duration of the eclipse of the centre of the primary Roche lobe in a
semi-detached binary yields a relation between two key parameters: {\it q}, 
the ratio of the mass of the secondary to that of the primary, and {\it i}, 
the inclination of the observer with respect to the binary orbital plane 
(Horne 1985). In the present situation, things are more complicated. 

The standard approach to measuring the duration of eclipse of the centre of
the primary Roche lobe is to measure the duration from mid-ingress to 
mid-egress of the white dwarf primary. This relies on detecting the contact 
points of eclipse of the white dwarf. Unfortunately, the observed eclipses 
are dominated by the ingresses and ingresses of the two spots. Can
the eclipse of the white dwarf nevertheless be detected? An estimate of 
the relative brightness of the white dwarf can be gauged from the hiatus 
between the disappearances of the spots on ingress (phases 0.970 to 0.975 
in Fig.~\ref{fig3}) or the corresponding feature on egress. At these phases 
(see Fig.~\ref{fig7}), the white dwarf photosphere is being covered/uncovered 
by the limb of the secondary (it is possible that other sources of light such
as the mass transfer stream close to the white dwarf are also being 
covered/uncovered). Fig.~\ref{fig3} shows that the light curve is relatively 
flat in these phase intervals so that the white dwarf photosphere contributes 
very little light compared to the spots. We attempted to detect the contact 
points of the white dwarf by looking for changes in slope in the light curve 
compared to the slope in the phase intervals 0.970 to 0.975 or 1.025 to 1.030. 
We were unable to convince ourselves that these contact points could be
identified. We used a light curve synthesis program to simulate eclipses
for a wide variety of spot locations and the results lead us to believe that 
the contact points of the white dwarf must be (almost) co-incident with the
contact points of the spots. However, it was difficult to explore all
possible combinations of parameters so we cannot prove this conjecture.

In the absence of detection of the white dwarf, it is impossible to define 
the times of eclipse of the centre of the Roche lobe of the white dwarf. 
The symmetry of the ingresses and egresses of the spots constrains where 
the spots can lie with respect to each other. More specifically, the very 
close match between the times of mid eclipse of both spots shows that the 
spot locations must lie on the same longitude in this specific reference
frame: its origin is at the binary centre of mass; it rotates at the binary
orbital period; its x-axis is defined by the line of sight from the earth
observer to the binary centre of mass at the time of mid-eclipse of both
spots; its x-y plane is tilted to the binary orbital plane by the system 
inclination. If this were not true, there would be an asymmetry in the 
timings of the ingresses and egresses of the spots. Longitudes in the
reference frame described above correspond to similar but not necessarily 
precisely equal longitudes in the reference frame of the white dwarf.
Such a result is what might be expected if the threading region gives rise 
to material moving to two spots which are located at the footprints on the 
white dwarf corresponding to the same magnetic field line, or to two field 
lines in close proximity.

There is, however, no constraint on the co-latitudes of the spots
(defined in the conventional manner as the angle between
the rotation axis of the white dwarf and the line connecting the
centre of the white dwarf and the accretion spot). Either spot might
lie at the rotation poles of the white dwarf, or at any co-latitude in
between, although, as we will now show, there is a constraint on the
{\it difference} of their co-latitudes; and even this difference depends 
on inclination and mass ratio. We can thus only define bounds for the 
allowable ({\it q, i}) combinations.

\begin{figure}
\begin{center}
\epsfxsize=7.8cm
\epsffile{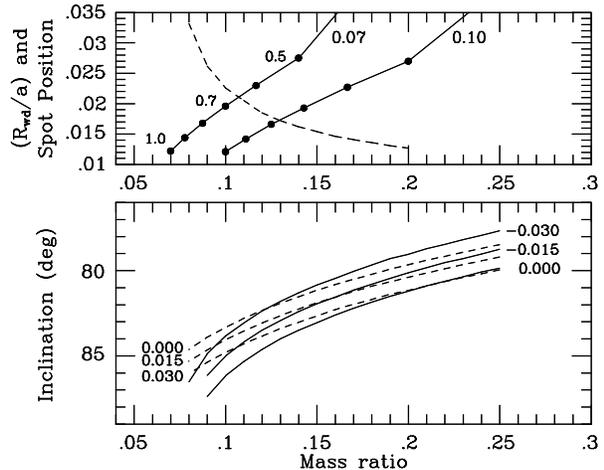}     
\caption{Lower panel: plot of inclination vs mass ratio. The continuous lines 
with negative numerical labels refer to the displacement of Spot 1 below
the orbital plane, the dashed lines with positive numerical labels refer to 
the displacement of Spot 2 above the orbital plane. See text for further
details. Upper panel: plot of distance of spots with respect to the orbital 
plane (dashed line) and radius of the white dwarf (continuous lines) vs
mass ratio. Labels 0.07 and 0.10 refer to the mass of the secondary, labels
0.5, 0.7 and 1.0 refer to the mass of the primary. See text for further 
details.}
\label{fig5}       
\end{center}
\end{figure}

The approach taken is to define one boundary by estimating the largest 
reasonable radius for a white dwarf in such a binary, relative to the binary
orbital separation, and place the accretion spots at this distance from the 
centre of the white dwarf above and below the orbital plane. To estimate the 
largest reasonable relative white dwarf radius, $R_{wd}/a$, the smallest 
plausible mass white dwarf in a binary with the smallest plausible total mass 
must be selected. The study of primary and secondary masses in cataclysmic 
variables by Smith \& Dhillon (1998) suggests minimum masses of 0.07 and 0.5 
M$_\odot$ for the secondary ($M_2$) and primary ($M_{wd}$) masses, respectively.
The white dwarf models of Wood (1995) along with Kepler's third law yield 
$R_{wd}$ of 10$^9$ cm and binary orbital separation, {\it a}, of $3.75 \times 
10^{10}$ cm, giving a relative radius for such a white dwarf of 0.027~{\it a}. 
Allowing for an additional 10 per cent if the spots are located above the 
surface of the white dwarf, we obtain 0.030~{\it a} for the largest relative 
displacement of the spots above and below the orbital plane. 

An eclipse simulation program similar to the kind described in section 2.1
of Horne (1985) to synthesize eclipse light curves of accretion discs was 
written. The program is functionally the same except that, in this 
case, eclipse light curves of white dwarfs and their accretion regions were 
modelled.  The number of free parameters was small: the mass ratio and 
inclination; the brightness of the white dwarf photosphere; the brightness,
location and size of the spots. Combinations of {\it q} and {\it i} were 
determined which gave eclipses of spots located as described in the previous 
paragraph and whose ingress/egress occurred at the observed orbital phases of 
$\pm0.0297$ (Spot 1) and $\pm0.0252$ (Spot 2). 

The other boundary was defined by determining combinations of {\it q} and {\it 
i} which gave eclipses of the centre of the white dwarf between -0.0297 and 
-0.0252 (i.e. one or other spot is co-incident with the centre of the white 
dwarf projected on the plane of the sky). An intermediate case of spots located
at 0.015~{\em a} was also computed. The results are shown in the lower panel
of Fig.~\ref{fig5}.

The continuous lines with negative numerical labels refer to Spot 1, the 
dashed lines with positive numerical labels refer to Spot 2. The numerical 
labels refer to the displacements, in units of {\em a}, of the relevant 
spot with respect to the centre of the white dwarf. Provided that the input 
assumptions are correct, allowable ({\it q}, {\it i}) combinations
lie somewhere in the locus defined by the intersection of the region
bounded by the -0.03~{\it a} and 0.0~{\it a} continuous lines and the 
region bounded by the 0.0~{\it a} and 0.03~{\it a} dashed lines. Once a
specific value of {\it q} and {\it i} is fixed, the {\it difference}
in co-latitudes between the spots is then also fixed.

\begin{figure}
\epsfxsize=7cm
\epsffile{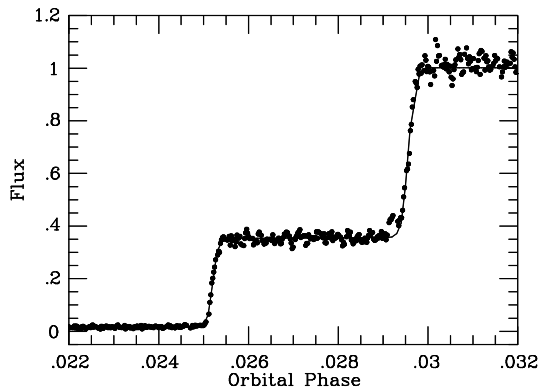}     
\caption{Eclipse egress of the second eclipse from the night of 2005 Sep
6/7 (points) and synthetic light curve using parameters described in the
text (continuous line). }
\label{fig6}       
\end{figure}

Fast photometric observations in the low state of the same kind as presented 
here will hopefully reveal the contact times of the white dwarf. With this 
information the exact value of {\it q} and {\it i} can be determined from the 
intersection of the appropriate continuous line for Spot 2 with the appropriate 
dashed line for Spot 1. For the moment, however, the necessary information is 
unavailable. Note that there is a minimum total separation of about 
$\sim0.017$~{\it a} (at $q\sim0.25$) for the spots: if the separation 
is smaller, there is no intersection of the region bounded by the dashed 
lines with the corresponding region bounded by the continuous lines. At
the lowest mass ratios, there is a maximum total separation of about 
$\sim0.06$~{\it a} (at $q\sim0.08$).

The upper panel of Fig.~\ref{fig5} enables further conclusions about the
allowed range of binary parameters to be drawn. The dashed line in
this panel was calculated under the assumption that the two spots are
equidistant from the centre of the white dwarf and the binary orbital 
plane. The locus of allowed ({\it q}, {\it i}) combinations is 
no longer a bounded area as in the lower panel, but a curve.  
This curve maps to the dashed line in the upper panel which shows as a 
function of {\it q} the distance of the spots from the orbital plane
in units of {\it a}. 

To explain the continuous lines in this panel, a consideration of the masses
of the component stars is needed. The revised orbital period-secondary
mass relation of Smith \& Dhillon (1998) predicts $M_2\sim$ 0.070 M$_\odot$
in SDSS J015543+002807 (averaging their linear and power
law fits). Older relationships (e.g. Warner 1995) expect 0.10 M$_\odot$.
The secondary in OY Car, an eclipsing dwarf nova with a very similar orbital
period, has a mass of 0.07 M$_\odot$ and this is likely to be accurate, 
as the component masses of the system were measured by one of the most 
reliable measurement techniques known (Wood et al. 1989). Thus, 0.07 M$_\odot$ 
is the best estimate for the mass of the secondary in SDSS J015543+002807.

As far as the primary is concerned, as noted above, a lower limit of 0.5
M$_\odot$ for the white dwarfs in cataclysmic variables is apparent in
fig. 5 of Smith \& Dhillon (1998).

The horizontal co-ordinate of the continuous lines in the upper panel is 
based on the mass ratio resulting from, respectively, $M_2 = 0.07$ or 
0.10 M$_\odot$ (large labels at upper right of each continuous line) 
combined with $M_{wd}$ ranging from 0.5 to 1.0 M$_\odot$: the filled 
circles on the continuous lines show $M_{wd}$ of 0.5, 0.6, 0.7, 0.8, 0.9 
and 1.0 M$_\odot$ with the 0.5, 0.7 and 1.0 points labelled. The vertical 
co-ordinate is both the relative distance of the spots above the orbital
plane, and the ratio of white dwarf radius to orbital separation, with the 
white dwarf radii taken from Wood (1995).

If the white dwarf mass is such that $R_{wd}/a$ 
lies {\em below} the dashed line, at least one of the spots
will lie above the white dwarf surface. It is not expected that accretion
columns will lie more than, say, 0.1 $R_{wd}$ above the white dwarf
surface, so the plausible range of $M_{wd}$ is 0.5 to 0.7 M$_\odot$ if 
$M_2$ has a mass of 0.07 M$_\odot$, or 0.5 to 0.8 if $M_2$ is 0.1 M$_\odot$. 
The same line of reasoning suggests an upper limit of 0.20 or 0.14 for {\it 
q}, again for $M_2 = 0.10$ or 0.07 M$_\odot$, respectively. Note that even 
if $R_{wd}$ is larger than the projected spot position, this does {\em 
not} imply that the accretion region lies on the photosphere of the white
dwarf; it is impossible from the present data to distinguish between this
and an accretion column seen projected on the photosphere.

In order to make further progress in the analysis, it is necessary to
adopt a specific choice of parameters. We choose M$_2$ = 0.070 M$_\odot$,
M$_1$ = 0.60 M$_\odot$, giving $q = 0.1167, a = 3.96 \times 10^{10}$ cm, 
and, from Wood (1995), $R_{wd} = 8.9 \times 10^8$ cm $= 0.0225\ a$. 
We urge that while these parameters are plausible given the constraints
described above, they should be regarded as a {\em choice} and not a 
determination, especially as it is likely that further observations will 
soon lead to much tighter estimates of these parameters.

\begin{figure*}
\begin{center}
$\begin{array}{c@{\hspace{1cm}}c}
\epsfxsize=6.5cm
\epsffile{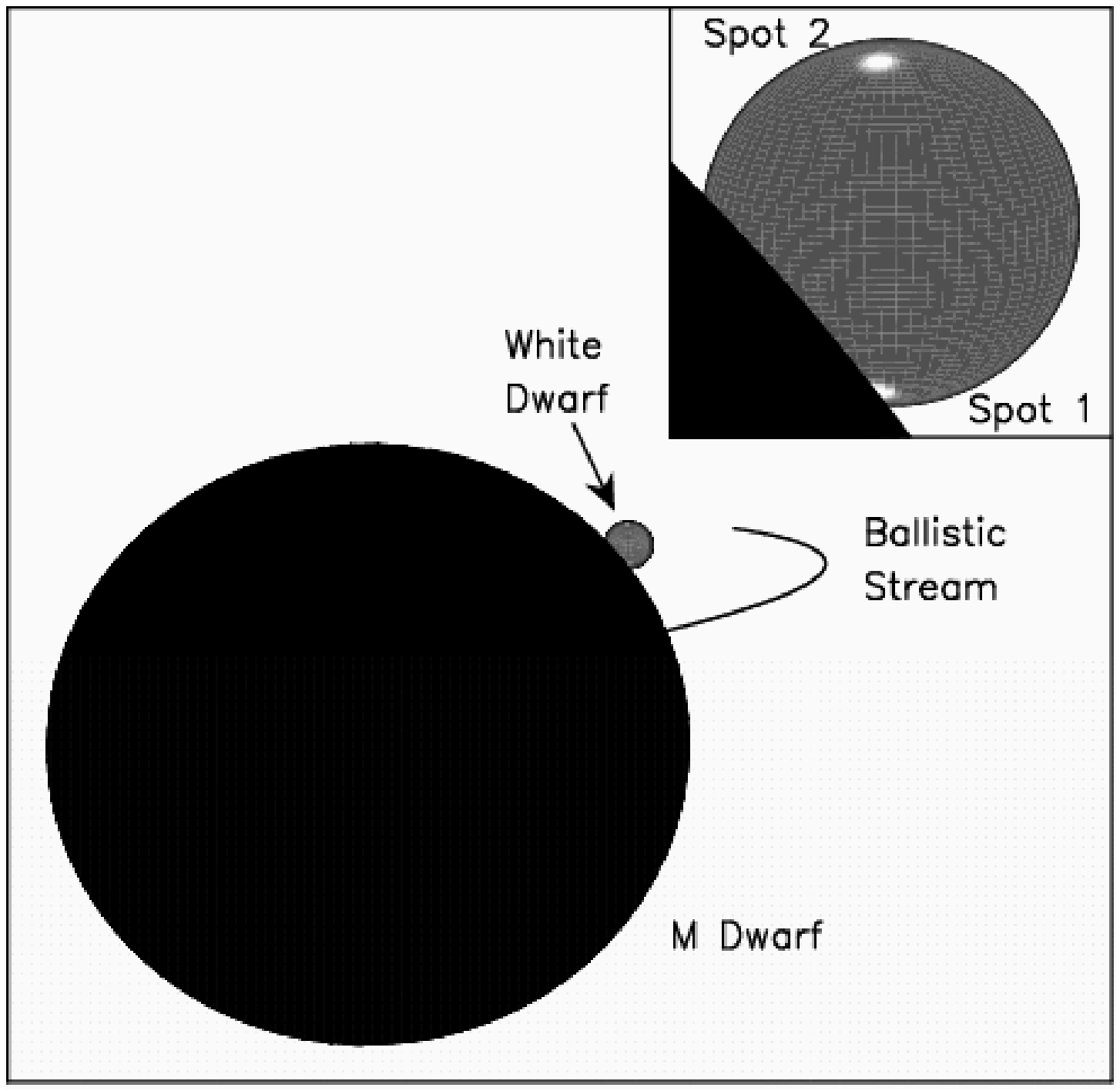} &
\epsfxsize=10cm
\epsffile{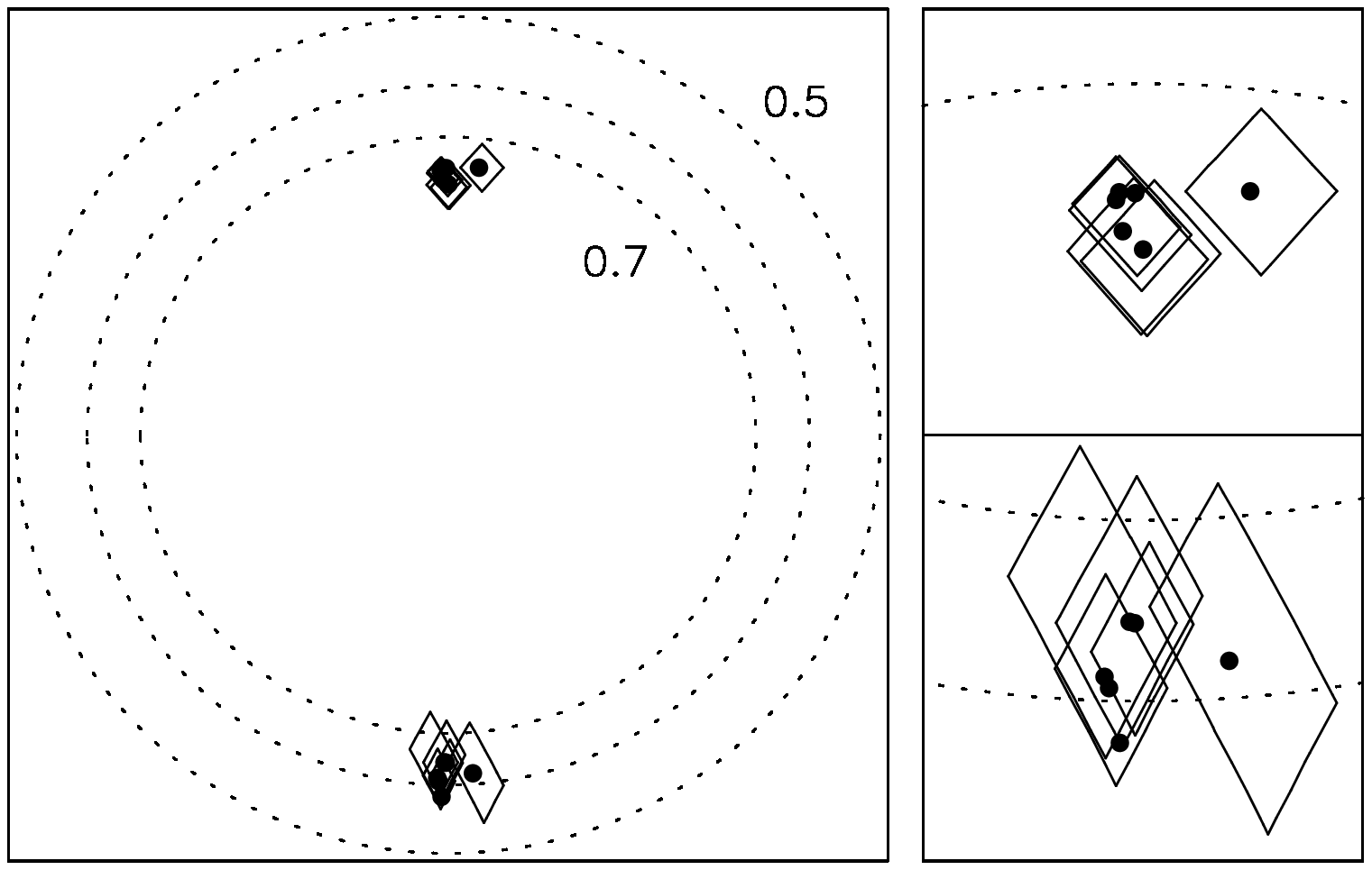} \\
\end{array}$
\end{center}
\caption{(Left) View of SDSS J015543+002807 at orbital phase 0.97. The
components of the system are labelled. The upper right inset shows the
view of the white dwarf at a larger scale.  (Right) Eclipse diamonds
defined by the limb of the secondary projected on the plane of the sky
at the phases of the four contacts of the eclipses of the spots. The
filled circles are the light centres of the spots as defined in the
text. The dotted lines show the outlines of white dwarfs with masses
of 0.5, 0.6 and 0.7 M$_\odot$. The size of the left panel is 0.058 $a$
(where $a$ is the binary orbital separation). The right panels show
the left panel at larger scale around the diamonds. The size of these
panels is 0.008 $a$. See text for more details. }
\label{fig7}       
\end{figure*}

\begin{table*}
\caption{Eclipse Measurements
{\label{tab:ecliparams}}}
\begin{tabular}{@{}cccccccccc}
       &    Date     &   First     &    Mid      &    Second   &    Third    &    Mid      &   Fourth  &  Ingress   &  Egress \\
       &             &  Contact    &  Ingress    &    Contact  &  Contact    &   Egress    &   Contact &  Duration  & Duration\\
       &    2005     &   (cyc)     &   (cyc)     &    (cyc)    &    (cyc)    &    (cyc)    &   (cyc)   &   (cyc)    &   (cyc) \\ 
       &             & $\pm0.0001$ & $\pm0.0001$ & $\pm0.0001$ & $\pm0.0001$ & $\pm0.0001$ &$\pm0.0001$&$\pm0.0001$ &$\pm0.0001$\\
\\
       & Aug 05/06   &             &             &             &   0.0296    &   0.0298    &  0.0301   &            &   0.0004  \\
       & Aug 06/07   &  -0.0301    &  -0.0299    &             &   0.0296    &   0.0297    &  0.0299   &            &   0.0003  \\
       & Aug 06/07   &  -0.0298    &  -0.0296    &  -0.0295    &   0.0295    &   0.0296    &  0.0298   &   0.0004   &   0.0003  \\
Spot 1 & Aug 10/11   &  -0.0299    &  -0.0297    &  -0.0294    &   0.0294    &   0.0296    &  0.0298   &   0.0005   &   0.0004  \\
       & Aug 11/12   &  -0.0299    &  -0.0297    &  -0.0295    &             &             &           &   0.0004   &           \\
       & Sep 06/07   &  -0.0300    &  -0.0298    &  -0.0296    &   0.0294    &   0.0296    &  0.0298   &   0.0003   &   0.0004  \\
       & Sep 06/07   &  -0.0299    &  -0.0298    &  -0.0295    &   0.0291    &   0.0296    &  0.0297   &   0.0004   &   0.0006  \\
       & Sep 07/08   &  -0.0296    &  -0.0294    &  -0.0292    &   0.0296    &   0.0299    &  0.0303   &   0.0004   &   0.0007  \\
\\
       & Aug 05/06   &  -0.0252    &   -0.0250   &   -0.0248   &             &             &           &   0.0004   &           \\
       & Aug 06/07   &  -0.0256    &   -0.0254   &   -0.0252   &    0.0251   &    0.0252   &  0.0256   &   0.0004   &  0.0004   \\
       & Aug 06/07   &             &   -0.0252   &   -0.0251   &    0.0250   &    0.0252   &  0.0254   &            &  0.0005   \\
Spot 2 & Aug 10/11   &  -0.0253    &   -0.0251   &   -0.0249   &             &             &           &   0.0004   &           \\
       & Aug 11/12   &  -0.0256    &   -0.0253   &   -0.0251   &    0.0252   &    0.0253   &  0.0256   &   0.0004   &  0.0004   \\
       & Sep 06/07   &  -0.0255    &   -0.0253   &   -0.0252   &    0.0250   &    0.0251   &  0.0255   &   0.0003   &  0.0004   \\
       & Sep 06/07   &  -0.0254    &   -0.0253   &   -0.0252   &    0.0250   &    0.0251   &  0.0254   &   0.0003   &  0.0004   \\
       & Sep 07/08   &  -0.0251    &   -0.0249   &   -0.0246   &    0.0253   &    0.0255   &  0.0258   &   0.0005   &  0.0005   \\
\\
       &             &   Spot 1    &   Spot 2    &    Spot 1   &    Spot 2   &  Delta  \\
       &             & Mid Eclipse & Mid Eclipse &  Duration   &  Duration   & Duration\\
       &             &   (cyc)     &    (cyc)    &     (cyc)   &     (cyc)   &   (cyc) \\
       &             & $\pm0.0001$ & $\pm0.0001$ & $\pm0.0001$ & $\pm0.0001$ & $\pm0.0001$ \\
\\
       & Aug 05/06 \\
       & Aug 06/07  &   -0.0001    &   -0.0001   &     0.0298  &    0.0253   &  0.0045 \\
       & Aug 06/07  &    0.0000    &    0.0000   &     0.0296  &    0.0252   &  0.0044 \\
       & Aug 10/11  &    0.0000    &             &     0.0296  &             &         \\
       & Aug 11/12  &              &             &             &    0.0253   &         \\
       & Sep 06/07  &   -0.0001    &   -0.0001   &     0.0297  &    0.0252   &  0.0045 \\
       & Sep 06/07  &   -0.0001    &   -0.0001   &     0.0297  &    0.0252   &  0.0045 \\
       & Sep 07/08  &    0.0003    &    0.0003   &     0.0297  &    0.0252   &  0.0045 \\
\end{tabular}
\end{table*}

With the above choices, the light curve synthesis program mentioned
previously was used to fix $i$ and locate the spots with 
respect to the white dwarf. The resulting values were $i = 83.5{\rm ^o}$, 
and co-latitudes $\beta = 125^{\rm o}$ and $25^{\rm o}$ for Spots 1 and 2,
respectively (although we must remind the reader that only the difference 
of the co-latitudes is constrained by our data). The longitude of each spot 
on the white dwarf was chosen to
be 0$^{\rm o}$ which is along the line joining the centres of the two stars.
The fit of the synthetic light curve to the egress of the second eclipse on 
2005 Sep 6/7 is shown in Fig.~\ref{fig6}. We re-iterate the previous caution 
that these parameters are only indicative and not determinations.

It is worth noting that Perryman et al (2001), in their analysis of
the similar features in the eclipse light curve of UZ For, derived a
separation in latitude of the two spots to be in the range $130 -
150^{\rm o}$.  Indeed, they go on to suggest that variations in this
separation may be a consequence of variations in the mass accretion
rate.

\section{Spot Properties From Details Of The Eclipses}

While the binary parameters cannot yet be fixed, properties of the
spots are measurable from the eclipse light curves.
Table~\ref{tab:ecliparams} lists the contact timings of 
Table~\ref{tab:observations} with respect to the ephemeris of Equation~1. The
measurement error indicated, 0.0001 cycles or 0.52 s, is quite conservative, 
so differences in measurements of the same parameter from eclipse to eclipse
of larger than, say, 0.0001 are certainly real. The table shows 
that the durations of the eclipses of the spots are constant but there is a 
little jitter in the times when they occur. This suggests that the colatitude 
of the spots is constant but there are small variations in the longitudes. 
A convenient way of visualizing some of the information in the table is with 
the aid of Fig~\ref{fig7}.

The middle panel of Fig. 7 shows the position of eclipse ``diamonds"
derived from the contact phases of the eclipses of the spots in all the 8
available eclipses. The four sides of the diamonds are defined by the 
projection of the limb of the secondary star on the plane of the sky at the 
four contact phases. The intention is to define bounds for the region emitting
the light originating from each spot. The filled circle in the middle
of the corresponding diamond shows the ``light centre" of each spot, defined 
as the intersection on the plane of the sky of the projection of the limb of
secondary at the phase of mid-ingress and mid-egress of the spot. The
right hand small panels show the region of the spots in the left hand
panel at a larger scale for clarity.

The dotted lines indicate the relative diameters of white dwarfs with masses 
of 0.5, 0.6 and 0.7 M$_\odot$ (assuming $M_2 = 0.070$ M$_\odot$). 
We re-iterate that the white dwarf contact points were {\em not} detected
so its location with respect to the diamonds is only indicative: the centre
of the white dwarf could be displaced in longitude or latitude with respect 
to the spots.

The rightmost upper and lower diamonds originate from the same
eclipse, from the night of 2005 Sep 7/8, whose timing has already been
noted to differ from the ephemeris by more than measurement
error. With this exception, the other diamonds have considerable
overlap with each other. There are, however, changes in size and also
in shape, especially in the diamonds in the lower panel which are
associated with Spot 1. The projected area of Spot 1 varies between
3.1 and $9.6 \times 10^{-6}\ a^2$, which is between 0.4 and 1.3 per
cent that of a 0.5 M$_\odot$ white dwarf, or between 0.8 and 2.4 per
cent that of a 0.7 M$_\odot$ white dwarf. This assumes that $M_2 =
0.07$ M$_\odot$. The projected area of Spot 2 varies between 2.2 and
$4.4 \times 10^{-6}\ {\rm a^2}$, which is between 0.3 and 0.6 per cent
that of a 0.5 M$_\odot$ white dwarf, or between 0.6 and 1.1 per cent
that of a 0.7 M$_\odot$ white dwarf. The actual surface area of the
white dwarf occupied by these spots will of course be larger than
this, because we are only measuring the projected area of the spots on
the sky, which will be smaller by a factor that depends on the binary
parameters, and could be as much as 8 times larger for our chosen parameters.
A model-dependent conclusion is that the spots subtend at the centre of 
the white dwarf angles of between 4 and $7^{\rm o}$, depending on the 
mass of the white dwarf.  These spot sizes are comparable to those estimated 
by Cropper \& Horne (1994) for ST LMi, although larger regions have been 
estimated for V347 Pav (Potter et al 2000).

As is evident in Fig~\ref{fig7}, the boundaries of the spots and their
light centres move around from eclipse to eclipse by typically half
the linear size of the diamond. While changes in spot positions have
been seen before in polars, particularly associated with accretion
state changes (see e.g. Romero-Colmenero et al, 2003), these are the
first data of sufficient quality to see such variations in spot
location within an essentially constant accretion state.  The diamonds
associated with Spot 2 (upper panel) are more tightly distributed,
symmetric in shape and smaller in extent than the diamonds associated
with Spot 1 which appear elongated along a line of constant
longitude. This elongation is not very sensitive to the co-latitude of
Spot 1. Even when this spot was modelled with a co-latitude as extreme
as $\sim90^{\rm o}$, its diamonds were distinctly more elongated than
those of Spot 2.

It is also important to emphasize that the diamonds are only boundaries to
the accretion spots and only constrain the distribution of enhanced 
brightness within so as to touch each side of the diamonds. The generally 
symmetrical shape of the diamonds associated with Spot 2, therefore, does  
not imply a symmetrical shape for the distribution of light in the spot,
which could be more elongated. We defer to a later paper the determination
of the surface brightness within the diamonds from the detailed shape of
the light curves during ingress/egress of the spots.

\section{Comparison with previous observations and discussion}

Both W2004 and Woudt et al. (2004) observed the system in its low state.
They report eclipse durations of 312 and 320 s, respectively, which is
consistent (given the lower time resolution and precision of their data)
with the duration of the eclipse of Spot 1. In the low state, therefore,
this remains the most luminous spot. In the polarimetry of W2004, positive
circular polarization is clearly visible around orbital phase 0.2-0.3.
This is consistent with spot 2 being visible for longer than spot 1,
and positive circular polarization coming from spot 2 with negative
circular polarization coming from spot 1.

The maximum in the orbital light curve, which occurs {\it after} the
eclipse, is (at least in part) the result of absorption of light which 
occurs {\it before} eclipse, presumably in the accretion flow/funnel 
reported by S2005 (and see below). Two further pieces of evidence favour 
this interpretation: (i) the appearance of the light curves shown in
fig. 15 of Woudt et al. (2004) and fig. 1 of W2004 which strongly
suggest that the bright phase of Spot 1, lasting from orbital phase
0.70 to 1.35, is ``cut into" by absorption; (ii) the centre of the
Spot 1 bright phase, $(0.7+1.35)/2$, is consistent with Spot 1 closest
to the line of sight at mid-eclipse. Consideration must also be
given to the contribution of spot 2 as well as variations caused by
cyclotron beaming from both spots.

An additional constraint on the position of Spot 2 might be obtained from 
noting that our eclipse light curves suggests that all but a few per cent 
of the optical light arises from the two spots. Around orbital phase 1.3, 
there is far more optical flux visible (Fig.~$\ref{fig1}$) than a few
per cent. On this basis, therefore, spot 2 never completely disappears from 
view, requiring that the sum of its co-latitude and the system inclination 
is less than 90$^{\rm o}$, which in turn suggests that its co-latitude is a 
few degrees or less. This constraint can be ``avoided" if, in the case
that spot 2 does disappear from view, some other light source in the
system increases in brightness and compensates for the loss from spot 2.
Heating of the inside face of the M star, or orbital phase dependent
variations in the apparent brightness of the accretion flow might behave
in this manner. On the other hand, fig. 1 of S2005 suggests that both spots 
do, in fact, disappear: around orbital phase 0.4, the (hard) x-ray flux
falls to zero, or almost zero.

In the low state, there is evidence that Spot 2 is still luminous: the right
hand eclipse in the lower panel of fig. 15 of Woudt et al. (2004), which was
obtained with 10 s time resolution, suggests a hiatus in between the 
eclipses of Spot 1 and Spot 2. This implies that the overall shape of the 
eclipse is the same as in our observations so that Spot 2 is still luminous.
The significant circular polarization apparent in fig. 1 of W2004 and
the substantial (and variable) amount of light visible at orbital phase 0.5
in fig. 15 of Woudt et al. (2004) also supports this interpretation. This
plot suggests that the relative brightnesses of Spots 1 and 2 are roughly the 
same as in the high state. 

The pre-eclipse dips apparent in Fig.~\ref{fig1} occur from at least
orbital phase 0.89 to 0.96. This is the same range of phases during which
S2005 found P Cyg-shaped spectral features which they attribute to 
absorption by the accretion flow. Their fig. 12 shows that the threading
region extends over approximately a factor of two in distance from the
white dwarf.

The time resolution of the observations presented here is sufficient to
warrant a search for quasi-periodic oscillations. In only one run, $\sim10$ 
cycles were seen with a period of about 2 s. However, a single detection 
made with equipment on a new telescope urges restraint from claiming that
the oscillation was intrinsic to the star.

\section{Summary and Conclusions}

SALT has been completed and first science observations of the polar SDSS
J015543+002807 have been obtained. These are high speed unfiltered photometric
observations with time resolution as short as 112 ms and are the highest 
quality observations of this kind of any polar to date. The target was observed 
in its high luminosity state and shows rapid eclipses of two compact regions
comprising unequivocal evidence for two accretion regions on or near the 
surface of the white dwarf. These account for all the optical light emitted 
by the system except for 1.5 per cent at mid eclipse and $\sim3-4$ per cent 
from the photosphere of the white dwarf (as distinct from the spots). 

The binary system parameters, $q$ and $i$ have been constrained. Further
limits, if M$_2 \sim 0.07$ M$_\odot$, are $0.5 < M_{wd} < 0.7\ M_\odot$,
$q < 0.14$; or, if M$_2 \sim 0.10$ M$_\odot$, $0.5 < M_{wd} < 0.8\ M_\odot$, 
$q < 0.20$. The correlation between $i$, $R_{wd}$ and the co-latitudes of 
the accretion regions has been explored, and the relative sizes and movements 
from eclipse to eclipse of the spots have been determined. 

In the absence of the detection of the white dwarf, there are few secure
constraints on the co-latitudes of the spots. The symmetry of the eclipses
of each spot suggest that they are located within a small range in
longitude.

The dominance of the accretion regions in the luminosity budget of SDSS 
J015543+002807, combined with its eclipses, offers unusual promise in
exploring the details of the accretion regions in a polar. Further
observations involving time-resolved spectrophotometry, polarimetry
and spectropolarimetry, all of which are available on SALT, 
are likely to uncover even more information. This potential can be fully 
exploited only when high time resolution observations during the low 
luminosity state of the system are obtained. It is likely that the contact 
points of the white dwarf will then be visible and this information will allow 
the position of the accretion regions with respect to the white dwarf to 
be fixed.

\section{Acknowledgments}

We are immensely indebted to the late Dr. Bob Stobie. Without his tireless
effort and enthusiasm in laying down the foundations of the SALT project
at its inception, the observations reported in this paper would never have 
been made. 

We are also immensely indebted to the technical staff of the HET for their 
generosity in sharing with us which aspects of the design of their telescope 
were successful and, especially, which were not.

We gratefully acknowledge the shareholders and directors of the SALT
Foundation for usage of the telescope: the National Research Foundation of
South Africa; Nicolaus Copernicus Astronomical Centre of the Polish
Academy of Sciences; the Hobby-Eberly Telescope Founding Institutions; 
Rutgers, the State University of New Jersey; Georg-August-Universitat
Goettingen; the University of Wisconsin at Madison; Carnegie-Mellon 
University; the University of Canterbury, New Zealand; the United
Kingdom SALT Consortium; the University of North Carolina - Chapel Hill;
and Dartmouth College. 

We thank an anonymous referee for a very thorough examination of the
paper leading to substantial improvements.

\end{document}